\newif\ifanonymous
\anonymousfalse   

\documentclass[orivec]{article}

\usepackage{xspace}
\usepackage{listings}
\usepackage{color}
\usepackage{xcolor}
\usepackage{adjustbox}
\usepackage{bold-extra}
\usepackage{comment}
\usepackage{caption}
\usepackage{paralist}
\usepackage{pifont}
\usepackage{wrapfig}
\usepackage{framed}

{
}

\lstdefinelanguage{XML}
{
  linewidth=0.97\textwidth,
  tabsize=3,
  frame=single,
  basicstyle=\scriptsize\ttfamily\color{black}\bfseries,
  rulesepcolor=\color{gray},
  xleftmargin=0pt,
  breaklines=true,
  showstringspaces=false,
  mathescape=true,
  escapeinside={(*}{*)}, 
  morestring=[b]",
  morecomment=[s]{<?}{?>},
  stringstyle=\color{green},
  keywords=[0]{eiserver,apps,app,appinfo,acronym,title,logo,
               desc,short,long,apphelp,content,execinfo,
               cmdlineapp,parameters,serverapp,selectone,selectmany,flag,hidden,permission,allowed,excluded,user,textfield,widgets,widget,widgetinfo,able,type_param,settings,eiout,eicommands,eiactions,eicommand,eiaction,line,lines,printonconsole,highlightlines,dialogbox,writefile,setcss,addmarker,addinlinemarker,oncodelineclick,onclick,cssproperty,cssproperties,option,default,elements,selector,examples,exset,file,folder,github,tagname,initialtext,ei\_response,ei\_server\_output,ei\_output,ei\_error,category,stream,download},
  keywordstyle=[0]\color{blue}\bf,
  keywords=[1]{id,visible,src,format,prefix,url,widgetid,name,max_size,method,check,value,version,from,to,fromch,toch,dest,autoclean,outclass,consoleid,consoletitle,color,filename,boxtitle,boxheight,boxwidth,path,repo,owner,branch,explicit,multiline,passinfile,selectable,icon,text,overwrite,trueval,falseval,execid,time},
  keywordstyle=[1]\color{orange},
  keywords=[2]{_ei_files,_ei_dirs,_ei_root,_ei_parameters,_ei_ppoints,_ei_outline,_ei_sessionid,_ei_clientid,_ei_execid,_ei_download,_ei_stream},
  keywordstyle=[2]\adjustbox{margin=0.4ex, bgcolor=red!20}
}
\lstdefinestyle{script}
{mathescape=false, numbers=left, stepnumber=1, numberstyle=\tiny, numbersep=10pt, language=bash, morekeywords={esac,in}}

\lstdefinestyle{shell}
{language=bash,keywords={}}

\let\lst\lstinline

\newcommand{\abs}[0]{\textsc{ABS}\xspace}
\newcommand{\ei}[0]{\textsc{EasyInterface}\xspace}

 \newcommand{\eigithub}[0]{\href{http://github.com/abstools/easyinterface}{\tt
     http://github.com/abstools/easyinterface}\xspace}
\newcommand{\envisage}{\textsf{Envisage}\xspace}
\newcommand{\envisagedesc}{{E}ngineering {V}irtualized {S}ervices}
\newcommand{\envisageurl}{\href{http://www.envisage-project.eu}{\tt http://www.envisage-project.eu}}

\newcommand{\applybutton}[0]{\begingroup\setlength{\fboxsep}{0pt}\colorbox{blue!20}{\texttt{Run}}\endgroup\xspace}
\newcommand{\refreshoutline}[0]{\begingroup\setlength{\fboxsep}{0pt}\colorbox{blue!20}{\texttt{Refresh Outline}}\endgroup\xspace}
\newcommand{\settingbutton}[0]{\begingroup\setlength{\fboxsep}{0pt}\colorbox{blue!20}{\texttt{Settings}}\endgroup\xspace}

\newcommand{\xmlstructref}[2]{\mbox{[\textcolor{red}{\uppercase{\small\ttfamily\bfseries #2}}]}}

\newfont{\mypfnt}{ptmri8t at 9pt}
\newcommand{\myparagraph}[1]{{\mypfnt #1}.\xspace}

\usepackage{hyperref}

\newcommand{\acks}[0]{%
  This work was partially funded by the EU project FP7-ICT-610582
  ENVISAGE: Engineering Virtualized Services, the Spanish MINECO
  projects TIN2012-38137 and TIN2015-69175-C4-2-R, and the CM project
  S2013/ICE-3006.}

\usepackage{authblk} 

\title{\ei: A toolkit for rapid development of GUIs for research
  prototype tools\thanks{\acks}  }

\author[1]{Jes\'us Dom\'enech}
\author[1]{Samir Genaim}
\author[2]{Einar Broch Johnsen}
\author[2]{Rudolf Schlatte}
\affil[1]{Complutense University of Madrid, Madrid, Spain}
\affil[2]{University of Oslo, Oslo, Norway}
\date{}

\ifanonymous
\author{Authors\inst{1}\inst{2}\institute{Institute 1 \and Institute 2}}
\renewcommand{\ei}[0]{\setlength{\fboxsep}{0pt}\colorbox{yellow}{\textsc{Anonymized}}\xspace}
\renewcommand{\envisage}[0]{\setlength{\fboxsep}{0pt}\colorbox{yellow}{\textsc{Anonymized}}\xspace}
\renewcommand{\abs}[0]{\setlength{\fboxsep}{0pt}\colorbox{yellow}{\textsc{Anonymized}}\xspace}
\renewcommand{\envisagedesc}[0]{\setlength{\fboxsep}{0pt}\colorbox{yellow}{\textsc{Anonymized}}\xspace}
\renewcommand{\envisageurl}[0]{\setlength{\fboxsep}{0pt}\colorbox{yellow}{\textsc{Anonymized}}\xspace}
\renewcommand{\eigithub}[0]{\url{http://github.com/}\setlength{\fboxsep}{0pt}\colorbox{yellow}{\textsc{Anonymized}}\xspace}
\renewcommand{\acks}[0]{%
  This work was partially funded by \setlength{\fboxsep}{0pt}\colorbox{yellow}{\textsc{Anonymized}}
}
\newcommand{\figsuf}[0]{-anonymized}
\else
\newcommand{\figsuf}[0]{}
\fi

\begin{document}

\maketitle
{
}

\begin{abstract}
%
  In this paper we describe \ei, an open-source toolkit for rapid
  development of web-based graphical user interfaces (GUIs).
  This toolkit addresses the need of researchers to make their research
  prototype tools available to the community, and integrating them in
  a common environment, rapidly and without being familiar with web
  programming or GUI libraries in general.
  If a tool can be executed from a command-line and its output goes to the
  standard output, then in few minutes one can make it accessible via a
  web-interface or within Eclipse.
  Moreover, the toolkit defines a text-based language that can be used
  to get more sophisticated GUIs, e.g., syntax highlighting, dialog
  boxes, user interactions, etc.
  \ei was originally developed for building a common frontend for
  tools developed in the \envisage~\cite{envisage} project.
 \end{abstract}


{
}

\section{Introduction}

During the lifetime of a research project, 
research prototype tools are often
developed which share many common aspects.
For example, in the \envisage \cite{envisage} project, we 
developed various tools for processing \abs programs: static
analyzers, compilers, simulators, etc.
%
Both as individual researchers and as groups,
we often develop several related tools over time to pursue a specific
line of research.

Providing the community with easy access to  research prototype tools 
is crucial to promote the
research, get feedback, and increase the tools' lifetime beyond the
duration of a specific project.
This can be achieved by building GUIs that facilitate trying tools;
in particular, 
tools with \emph{web-interfaces} can be tried without the overhead of first
downloading and installing the tools.

In practice, we typically avoid developing GUIs until tools are fairly
stable.  Since prototype tools change continuously, in
particular during a research project, they will often not be
available to the community during early development.
Both programming plug-ins for sophisticated frameworks such as Eclipse
Scout and building simpler GUIs from scratch are tedious tasks, in
particular for web-interfaces. It typically gets low priority when
developing a research prototype. Often we opt for copying the GUI of
one tool and modifying it to fit the needs of a new related tool.
Apart from code duplication, these tools will ``live'' separately,
although we might benefit from having them in a common GUI.

\ei is a toolkit that aims at simplifying the process of building and
maintaining GUIs for (but not limited to) research prototype
tools. Avoiding complex programming, it provides an easy, declarative
way to make existing (command-line) tools available via different
environments such as a web-interface, within Eclipse, etc.
It also defines a text-based output language that can be used to
improve the way results are presented to the user without requiring
any particular knowledge of GUI/Web programming; e.g., if the output
of a tool is (a structured version of) ``\emph{highlight line number
  10 of file ex.c}'' and ``\emph{when the user clicks on line 10, open
  a dialog box with the text ...}'', the web-interface will interpret
this and convert it to corresponding visual effects. An advantage of
using such an output language is that it will be understood by
all the front\-end environments of \ei, e.g., the web-interface and the Eclipse
plug-in (which is still under development).
\ei is  open source and available at \eigithub. 
A more detailed descreption on how to use \ei is availale in
Appendix~\ref{sec:methodology}, and in the user manual~\cite{ei}.
An online demo is available at \url{https://youtu.be/YE7YwR2duzk}.

{
}

\section{General Overview}
\label{sec:overview}

\begin{figure}[t]
\centering{\fbox{\includegraphics[width=\linewidth]{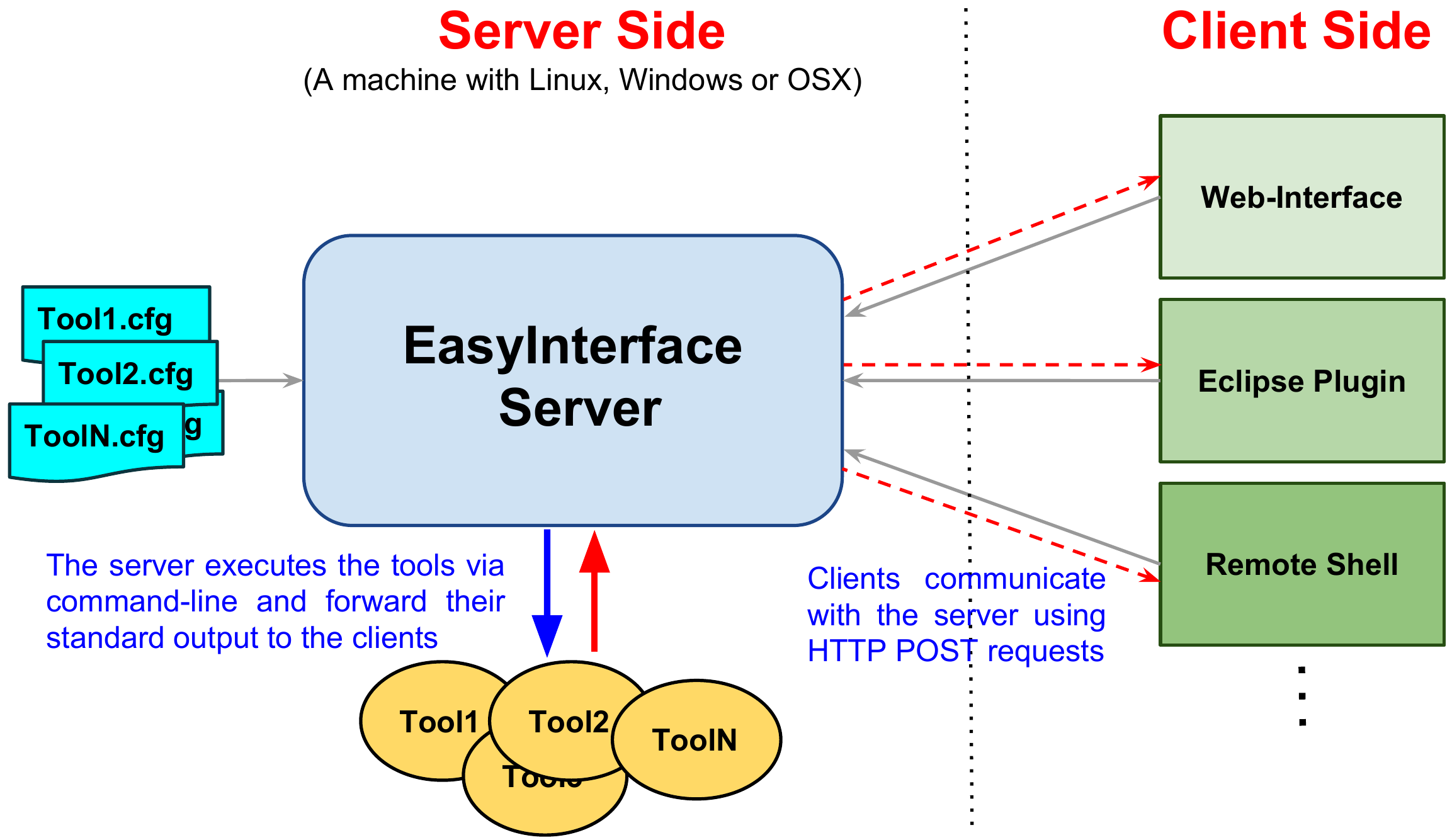}}}\\
\caption{\label{fig:eiframework} \ei architecture.}
\end{figure}

The overall architecture of \ei is depicted in
Fig.~\ref{fig:eiframework}. Its two main 
components are
\begin{inparaenum}[\upshape(\itshape i\upshape)]
\item \emph{server side}: a machine with several tools (the circles
  \texttt{Tool1}, etc.) executable from the command-line,
  and with output going to the standard output.  These are the
  tools that we want to make available for the outside world; and
\item \emph{client side}: several clients that 
  communicate with the server to execute a tool. Tools may run
  on the server machine or on other machines; e.g., the
  web-interface can be installed as a web-page on the server, 
  and accessed from anywhere with a web browser.
\end{inparaenum}
Clients can connect to several servers simultaneously.

The  server side addresses  the problem of
\emph{providing a uniform way to remotely execute locally installed tools}.
This problem is solved by the server, which consists
%
%
of PHP programs (on top of an HTTP server).  The server
supports \emph{declarative specifications} of how local tools can be
executed and which parameters they take, using simple configuration
files. For example, the following XML snippet is
a
configuration file for a tool called \lst{"myapp"}.  
\begin{center}
\begin{minipage}{0.9\textwidth}
\begin{lstlisting}
<app id="myapp" visible="true">
 ...
 <execinfo method="cmdline">
  <cmdlineapp>
    (*/path-to/myapp*) _ei_parameters
  </cmdlineapp>
 </execinfo>
 <parameters prefix = "-" check="false">
  ...
  <selectone name="c">
   <option value="1" />
   <option value="2" />
  </selectone>
 </parameters>
</app>
\end{lstlisting}
\end{minipage}
\end{center}
The
\lst{cmdlineapp} tag is a template describing how to execute the
tool from the command-line. 
The template parameter \lst{_ei_parameters} is replaced by an
appropriate value before execution.
The server also supports 
template parameters for, e.g.,
passing files, temporal working directories, session
identifiers, etc.
The \lst{parameters} tag includes a list of parameters accepted by the
tool. For example, the parameter ``\texttt{c}'' above takes one of the
values $1$ or $2$.

Once the configuration file is installed on the server, we can access
the tool using an HTTP POST request that includes JSON-formatted data
like the following one  
\begin{center}
\begin{minipage}{0.9\textwidth}
\begin{lstlisting}
{
 (*command: "execute",*)
 (*app\_id: "myapp",*)
 (*parameters:*) {
    (*c: ["1"],*)
    (*...*)
 },
 (*...*)
}
\end{lstlisting} 
\end{minipage}
\end{center}
When receiving such a request,
the server  generates a 
shell command according to the specification in the configuration file
(e.g., ``\texttt{/path-to/myapp -c 1}''), executes it and redirects
the standard output to the 
client.
The server also supports 
%
\begin{inparaenum}[\upshape(\itshape i\upshape)]
\item tools that generate output in the background, we let clients 
  fetch output (partially) when it is ready; and
\item tools that generate files, we let clients download them later
  when needed.
\end{inparaenum}
In all cases, the server can\emph{ restrict
the resources }
available to a tool (e.g., the processing time), and \emph{guarantees
  the safety} of the generated command;
i.e., clients cannot manipulate the
server to execute other programs installed on the server.
In addition to tools, 
the server can include example files, so users can easily try the tools.

\begin{figure*}[t]
\centering{\fbox{\includegraphics[width=\linewidth]{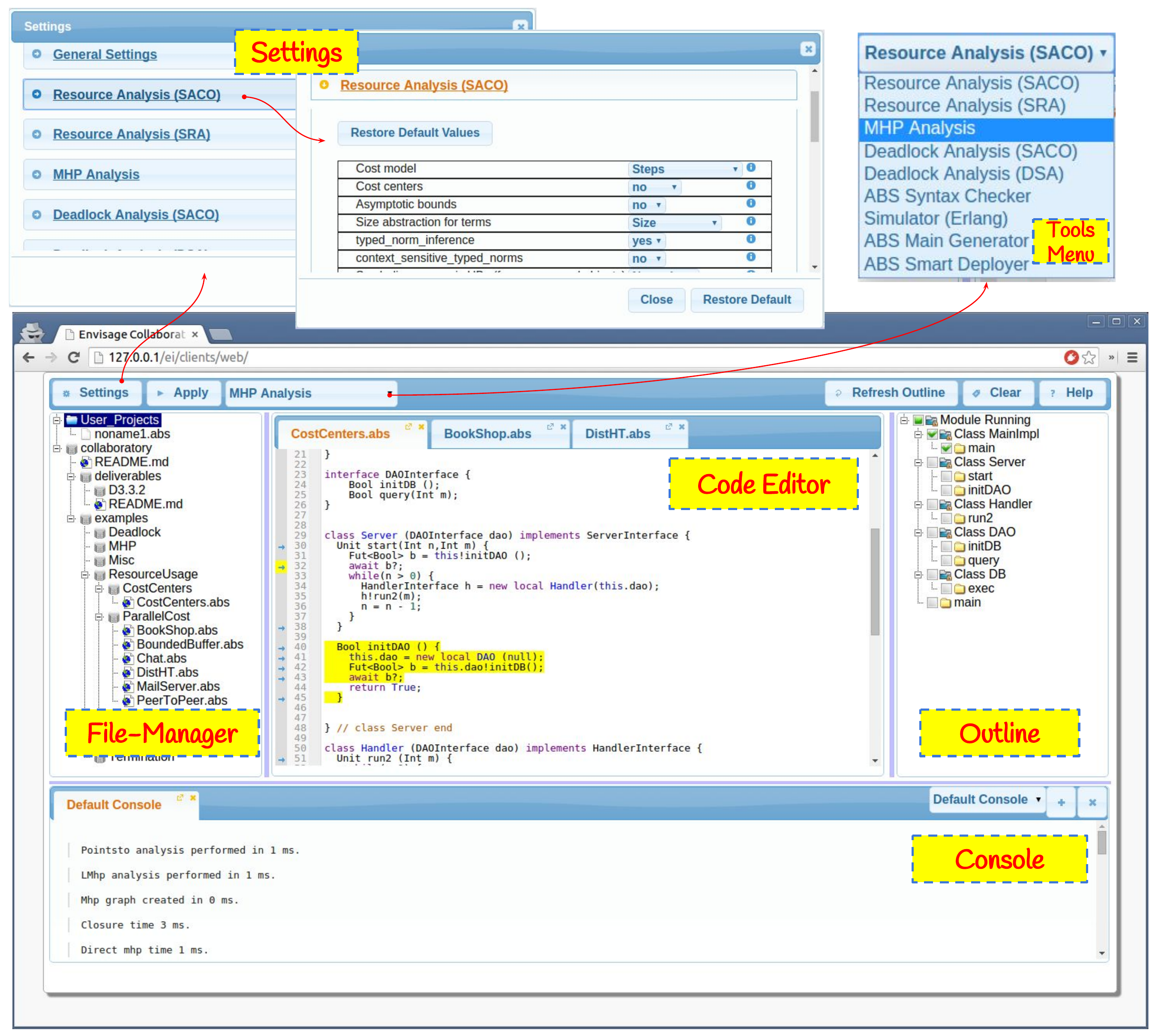}}}
\caption{\label{fig:webclient}\ei Web-Interface Client}
\end{figure*}

\ei not only makes the server side execution of a tool easy,
it provides client side GUIs that
\begin{inparaenum}[(1)]
\item connect to the server and ask for available tools;
\item let users select the tool to execute, set its parameters and
 provide a source program;
\item generate and send a request to the server; and
\item display the returned output.
\end{inparaenum}
\ei provides three generic clients: a \emph{web-interface} similar to an
IDE; 
an Eclipse IDE plug-in; and a remote command-line
shell.
The last two clients are under development, so we focus here on the
web-interface.

The web-interface, shown in Fig.~\ref{fig:webclient}, is
designed like an IDE where users can edit programs, etc. Next to the
\applybutton button there is a drop-down menu with
all available tools
obtained from the associated servers. In the settings window, the user
can select values for the different parameters of
each tool. These parameters are specified in the corresponding
configuration files on the server side, and automatically converted to
combo-boxes, etc., by the web-interface.
When the user clicks the \applybutton button, the web-interface sends
a request to the associated server to execute the selected tool and
prints the received output back in the console area of the
web-interface.

Since the web-interface and Eclipse plug-in are GUI based clients,
\ei allows tools to generate output with some
graphical effects, such as opening dialog-boxes, highlighting code lines,
adding markers, etc. To use this feature, tools
must support the \ei output language, as in the
following XML snippet
\begin{center}
\begin{minipage}{0.9\textwidth}
\begin{lstlisting}
<highlightlines dest="/path-to/sum.c"> 
  <lines> <line from="5" to="10"/> </lines>
</highlightlines>
...
<oncodelineclick dest="/path-to/sum.c" outclass="info">
  <lines><line from="17" /></lines>
  <eicommands>
    <dialogbox boxtitle="Hey!"> 
      <content format="text">(* some message *)</content>
    </dialogbox>
  </eicommands>
</oncodelineclick>
\end{lstlisting}
\end{minipage}
\end{center}
%
The tag \lst{highlightlines} indicates that Lines 5--10 of
file \texttt{/path-to/sum.c} should be highlighted. The tag
\lst{oncodelineclick} indicates that when clicking on Line $17$, a
dialog-box with a corresponding message should be opened.
Note that a tool is only modified once to produce such output, with
similar effect in all \ei clients (including future ones).

{
}

\section{Concluding Remarks}
\label{section:conclusions}

\ei is a toolkit for the rapid development of GUIs for command-line
research prototype tools. The toolkit has been successfully used in
the \envisage project to integrate the tools from the
different partners in a common web-based environment, including
parsers, type-checkers, compilers, simulators, deadlock
and worst-case cost analyzers, and a systematic testing framework (see \url{http://abs-models.org}).
Our experience suggests that the methodology implied by \ei for building GUIs is
adequate for research prototype tools; as  such tools
change continuously, the
corresponding GUIs can be modified immediately and with negligible effort.
Future work includes plans to develop more clients, and 
libraries for different programming languages to facilitate 
generation of the output commands/actions instead of
printing these directly.

\bibliographystyle{abbrv}

\newpage
\appendix

{
}

\section{Using \ei}
\label{sec:methodology}

In this appendix we describe the methodology that \ei implies for its
users by explaining the basics of its different parts and the way they
are supposed to be used.
In Sec.~\ref{sec:methodology:server} we describe the different options
that are provided by the \ei server, in
Sec.~\ref{sec:methodology:client} we review the different features
of the web-interface, and in Sec.~\ref{sec:methodology:eiout} we
discuss the \ei output language. 
Note that the installation of \ei is practically immediate: It only
consists in cloning the github repository and making its root
directory accessible via an HTTP server. Thus, we omit the
installation details here and refer the reader to the installation guide
that is available in the github repository for further information.

{
}

\subsection{Adding Tools and Examples to the Server}
\label{sec:methodology:server}

As described in Sec.~\ref{sec:overview}, adding a tool to the server
is simply done by providing a configuration file such as the one in
Sec.~\ref{sec:overview}.
This file includes two main components: (1) the command-line template
that describes how to execute the corresponding tool; and (2) the
parameter section that describes which command-line parameters the
tool will take.

\ei supports several types of parameters such as a parameter with a
single value, a parameter with multiple values, a Boolean parameter,
etc. Each parameter has a name, a set of valid values, and a set of
default values. When receiving a request to execute a tool (such as the
JSON-formatted data in Sec.~\ref{sec:overview}), the server generates
a sequence of command-line parameters from those specified in the
request and replaces the template parameter \lst{_ei_parameters} by
this list.
For example, if there is a parameter named ``\texttt{c}'' and its provided
value is ``\texttt{1}'', then what is passed to the tool is
``{$\mbox{-}$c
  1}''.
The prefix ``-'' that is attached to the parameter name can be
specified using the \lst{prefix} attribute in the parameters
section. In addition, the \lst{check} attribute indicates if the
server should reject requests with invalid parameter values.
Apart from \lst{_ei_parameters} the command-line template might
include other template parameters, all are first replaced by
corresponding values and then the resulting shell command is executed
and its output is redirected back to the client. Next we describe some
of the available template parameters:

\defaultleftmargin{1em}{}{}{}
\begin{compactitem}

\item \lst{_ei_files}: tools typically receive input files (e.g.,
  programs or models) to process. The execution request (i.e., the
  JSON-formatted data of Sec.~\ref{sec:overview}) can include such
  files, and, in order to pass them to the corresponding tool, the
  server first saves them locally in a temporary directory and replaces
  \lst{_ei_files} by a list of corresponding local paths.
  
\item \lst{_ei_outline}: since \ei was first developed for tools that
  process programs (e.g., program analysis tools), the execution
  request can include so-called \emph{outline entities}. These are
  elements of the input programs such as method names, class names,
  etc., and they are used to, e.g., indicate from where the tool
  should start the analysis. The server replaces \lst{_ei_outline} by
  the list of the provided outline entities.

\item \lst{_ei_execid}: this corresponds to a unique
  \emph{execution identifier} that is assigned (by the server) to the
  current execution request, which can be used for future references
  to this execution as discussed below.

\item \lst{_ei_stream}: there are tools that do not generate output
  immediately, such as simulators. In this case we would like to keep
  the tools running in the background and fetch their output
  periodically without maintaining the current connection to the
  server. The template parameter \lst{_ei_stream} corresponds to a
  temporary directory where the tool can write its output and where
  clients can fetch this output by corresponding requests to the
  server. These requests should include the corresponding
  \emph{execution identifier}. For example, the tool could output the
  following command (in the \ei output language) and terminate, while
  keeping a process running in the background (complying with the
  server's permissions) writing its output chunks to the
  \lst{_ei_stream} directory:
\begin{center}
\begin{minipage}{0.95\textwidth}
\begin{lstlisting}
<content format="text" execid="EI65231" time="60sec">>
  The program is running in the background, 
  the output goes here ...
</content>
\end{lstlisting}
\end{minipage}
\end{center}
This command indicates that the output, in \texttt{text} format,
should be fetched every 60 seconds using the execution identifier
specified in \lst{execid}. Note that it is the responsibility of the
client (e.g., the web-interface) to fetch the output once it receives
the above command.

\item \lst{_ei_download}: some tools generate output in the form of
  large files, e.g., compiled code. Instead of redirecting this output
  directly to the client it can be more convenient to return download
  links. This template parameter corresponds to a temporary directory
  where such files can be stored and later fetched by sending a
  special request to the server with the file name and the
  corresponding \emph{execution identifier}, similar to the stream
  mode above. The \ei output language includes a special command for
  downloading such files:
\begin{center}
\begin{minipage}{0.95\textwidth}
\begin{lstlisting}
<download  execid="EI65231" filename="file.zip" />
\end{lstlisting}
\end{minipage}
\end{center}
Once the client (e.g., the web-interface) receives this command, it
sends a request to download the file \texttt{file.zip} that is
associated with the execution identifier ``\texttt{EI65231}''.
The command can also be assigned to an ``on click'' action instead of
downloading it immediately (see Sec.~\ref{sec:methodology:eiout}).

\item \lst{_ei_sessionid}: this corresponds to a unique
  session identifier that is assigned to the user, and can be used to
  track the user's activity among several requests. It is implemented
  using PHP sessions.

\item \lst{_ei_clientid}: this corresponds to the client identifier,
  e.g., \texttt{webclient}, \texttt{eclipse}, etc. It can be used to
  generate client directed output which uses the \ei output language
  for the selected clients and plain text for others.

\end{compactitem}

\medskip
\noindent
The server configuration files also include options for controlling
security issues, timeout for tools, etc.

In addition to the tools, one can install example sets on the server
side, which are meant to be used by clients (e.g., the web-interface)
to provide users with some examples on which they can apply the
available tools. Adding example sets is done via configuration files
such as the following:
\begin{center}
\begin{minipage}{0.95\textwidth}
\begin{lstlisting}
<examples>
 <exset id="iter">
  <folder name="Examples_1">
   <folder name="iterative">
    <file name="sum.c" url="https://.../sum.c" />
    ...
   </folder>
   ...
  </folder>
 </exset>
 <exset id="set2">
  <folder name="Examples_2">
   <github owner="username" repo="examples" branch="master" path="benchmarks"/>
 </folder>
</exset>
</examples>
\end{lstlisting}
\end{minipage}
\end{center}
This defines two sets of examples. The first uses a directory
structure, while the second refers to a github repository.  Note that
in the first case the example files are not necessarily installed on
the server; a link must be provided for each file.

{
}

\subsection{Using the Web-Interface Client}
\label{sec:methodology:client}

The web-interface client of \ei is a JavaScript program that runs in a
web browser, a screenshot is depicted in Fig.~\ref{fig:webclient}. It
uses JQuery~\cite{jquery} as well as some other libraries like
JSTree~\cite{jstree} and CodeMirror~\cite{codemirror}.
It is designed like an IDE, this is because \ei was first developed
for integrating the tools of the \envisage~\cite{envisage} project
(static analyzers, simulators, etc.)  in a common GUI.
It includes the following components:
\begin{inparaenum}[(1)]
\item the \emph{code editor}, where users can edit programs; 
\item the \emph{file-manager} that contains a list of predefined
  examples and user files;
\item the \emph{outline view} that includes an outline (e.g., methods
  and classes) of one or more files;
\item the \emph{console} where the results of executing a tool is
  printed by default; and
\item the \emph{tool bar} that includes several buttons to execute a
  tool, etc.
\end{inparaenum}
Next we describe the workflow, and give more details on these
components.

\medskip
\noindent
\myparagraph{The Workflow}
The workflow within the web-interface is as follows:
\begin{inparaenum}[(a)]
\item write a new program or open an existing one from the
  file-manager;
\item click on the \refreshoutline button to generate the outline of
  the currently open program, and select some entities from this
  outline;
\item select a tool from the tools menu; and
\item click on the \settingbutton button to set the values of the
  different parameters;
\item click on the \applybutton button to execute the selected tool.
\end{inparaenum}
At this point the web-interface sends a request to the corresponding
server to execute the selected tool (passing the currently opened
file, parameter values, selected outline entities, etc. to the tool),
and the output is returned and printed in the console area of the
web-interface. If the tool's output is in the \ei output language,
then it passes through an interpreter that converts it to some
corresponding graphical output. The user can apply a tool (and
generate an outline) on several files by selecting the corresponding
option from the context menu in the file-manager (opened with a right
click on an entry in the file-manager).

\medskip
\noindent
\myparagraph{Code Editor} 
The code editor is based on CodeMirror~\cite{codemirror}, it can be
easily configured to use syntax-highlighting for different languages.

\medskip
\noindent
\myparagraph{Tools Menu}
The tools menu includes a list of tools that can be executed by the
user. This list can be set in the web-interface configuration file,
simply by providing the URLs of the corresponding \ei servers and, for each,
indicating if all available tools should be included or only some
selected ones. The default configuration of the web-interfaces fetches
all tools that are installed on the server running on the machine
where the web-interface is installed.

\medskip
\noindent
\myparagraph{Settings}
When clicking the \settingbutton button, a settings window is opened
where the user can choose values for the different parameters of the
different tools (see the top part of Fig.~\ref{fig:webclient}). This
is automatically generated using the parameters defined in the server
configuration file (the web-interface fetches this information from
the corresponding server).

\medskip
\noindent
\myparagraph{File-Manager}
The predefined examples included in the file-manager can be set in the
web-interface configuration file by providing the URLs of the
corresponding \ei servers and identifiers for the example sets to be
included. As we have seen in Sec.~\ref{sec:overview}, such a set can
simply be
given as a reference to a github repository. The file-manager
can also allow users to create their own files, upload files from local
storage, and clone public and private github repositories.

\medskip
\noindent
\myparagraph{Outline} 
The Outline area is supposed to include an outline of some of the
program files (available in the file-manager), and thus depends on
the structure of those programs. For example, in the case of the
\envisage~\cite{envisage} project it includes elements (methods,
classes, etc.) of \abs programs. Using it for Java programs, for
example, would require changing the way the outline is generated.  \ei
already provides an easy way to change the outline generator.
All we have to do is 
\begin{inparaenum}[(1)]
\item to provide a tool (installed on an \ei server, like any other
  tool) that takes a set of files and generates an XML structure that
  represents an outline, the web-interface will convert this XML to a
  tree-view; and 
\item to modify the web-interface configuration file in order to use
  this tool for generating the outline.
\end{inparaenum}

{
}

\subsection{Using The Output Language}
\label{sec:methodology:eiout}

The \ei output language is a text-based language that allows tools to
generate more sophisticated output. It is supported in both the
web-interface and  Eclipse clients. In this section we will explain
the basics of this language by example.
An output in the \ei output language is an XML of the following form:
\begin{center}
\begin{minipage}{0.95\textwidth}
\begin{lstlisting}
<eiout> 
 <eicommands>
    $\xmlstructref{eiout}{eicommand}$*
 </eicommands>
 <eiactions>
    $\xmlstructref{eiout}{eiaction}$*
 </eiactions>
</eiout>
\end{lstlisting}
\end{minipage}
\end{center}
where \xmlstructref{eiout}{eicommand}* is a list of commands to be
executed; and \xmlstructref{eiout}{eiaction}* is a list of actions to
be declared.
An \xmlstructref{eiout}{eicommand} is an instruction like: \emph{print
  a text on the console}, \emph{highlight lines 5-10}, etc.
An \xmlstructref{eiout}{eiaction} is an instruction like: \emph{when the
  user clicks on line 13, highlight lines 20-25}, etc.
In the rest of this section we consider some representative commands and
actions.

\medskip
\noindent
\myparagraph{Printing in the Console}
Recall that when the \ei output language is used, the web-interface
does not redirect the output to the console area and thus we need a
command to print in the console area.
The following is an example of a command that prints ``Hello World''
in the console area:
\begin{center}
\begin{minipage}{0.95\textwidth}
\begin{lstlisting}
<printonconsole consoleid="1" consoletitle="Welcome">
  <content format="text">Hello World</content>
</printonconsole>
\end{lstlisting}
\end{minipage}
\end{center}
The value of the \lst{consoleid} attribute is the console identifier
in which the given text should be printed (e.g., in the web-interface
the console area has several tabs, so the identifier refers to one of
those tabs). If a console with the provided identifier does not exist
yet, a new one is created, with a title as specified in
\lst{consoletitle}. If \lst{consoleid} is not given the output goes to
the default console.
Inside \lst{printonconsole} we can have several \lst{content} tags
which include the content to be printed. The attribute \lst{format}
indicates the format of the content. In the above example it is plain
\texttt{text}, other formats are supported as well, e.g.,
\texttt{html}, \texttt{svg}, and \texttt{graphs}. The \texttt{graphs}
option refers to diagrams that are drawn using DyGraphs
\cite{dygraphs}, where the data is provided inside the \lst{content}
tag using a JSON based format.

\medskip
\noindent
\myparagraph{Adding Markers}
The following is an example of a command that adds a marker next to a
code line in the editor area:
\begin{center}
\begin{minipage}{0.95\textwidth}
\begin{lstlisting}
<addmarker dest="path" outclass="info">
  <lines> <line from="4" /> </lines>
  <content format='text'> (*some associated text*) </content>
</addmarker>
\end{lstlisting}
\end{minipage}
\end{center}
The attribute \lst{dest} indicates the \emph{full path} of the file in
which the marker should be added.
The attribute \lst{outclass} indicates the nature of the marker, which
can be \texttt{info}, \texttt{error}, or \texttt{warning}. This value
typically affects the type/color of the icon to be used for the
marker.
The tag \lst{lines} includes the lines in which markers should be
added, each line is given using the tag \lst{line} where the \lst{from}
attribute is the line number~(\lst{line} can be used to define a
region in other commands, this is why the attribute is called
\lst{from}).
The text inside the \lst{content} tag is associated to the marker (as
a tooltip, a dialog box, etc., depending on the client).

\medskip
\noindent
\myparagraph{Highlighting Code Lines}
The following command can be used to highlight code lines:
\begin{center}
\begin{minipage}{0.95\textwidth}
\begin{lstlisting}
<highlightlines dest="path" outclass="info" > 
  <lines> <line from="5" to="10"/> </lines>
</highlightlines>
\end{lstlisting}
\end{minipage}
\end{center}
Attributes \lst{dest} and \lst{outclass} are as in the \lst{addmarker}
command. Each \lst{line} tag defines a region to be highlighted. In
the example above, lines 5--10 get highlighted. We can also use the
attributes \lst{fromch} and \lst{toch} to indicate the columns in
which the highlight starts and ends respectively.

\medskip
\noindent
\myparagraph{Opening a Dialog Box}
The following command can be used to open a dialog box with some
content:
\begin{center}
\begin{minipage}{0.95\textwidth}
\begin{lstlisting}
<dialogbox outclass="info" boxtitle="Some Title!" boxwidth="100" boxheight="100"> 
  <content format="html"> (* some associated text *) </content>
</dialogbox>
\end{lstlisting}
\end{minipage}
\end{center}
The dialog box will be titled as specified in \lst{boxtitle} and it
will include the content as specified in the \lst{content}
environments.

\medskip
\noindent
\myparagraph{CodeLine Actions}
A \emph{CodeLine action} defines a list of commands to be executed
when the user clicks on a line of code (more precisely, on a marker
placed next to the line). The commands can be any of those seen
above. The following is an example of such an action:
\begin{center}
\begin{minipage}{0.95\textwidth}
\begin{lstlisting}
<oncodelineclick dest="path" outclass="info" >
  <lines> <line from="17" /> </lines>
  <eicommands>
    <highlightlines>
      <lines> <line from="17" to="19"/> </lines>
    </highlightlines>
    <dialogbox boxtitle="Hey!"> 
      <content format="html"> (* some text *) </content>
    </dialogbox>
  </eicommands>
</oncodelineclick>
\end{lstlisting}
\end{minipage}
\end{center}
First note that the  XML description above should be placed inside the
\lst{eiactions} tag.
When the above action is executed, e.g., by the web-interface client,
a marker will be shown next to line 17 of the file specified in the
attribute \lst{dest}.
If the user clicks on this marker the commands inside the
\lst{eicommands} tag will be executed, and if the user clicks again
the effect of these commands is undone.
In the case above, a click highlights lines 17--19 and opens a dialog
box, and another click removes the highlights and closes the dialog
box.

\medskip
\noindent
\myparagraph{OnClick Actions}
OnClick actions are similar to CodeLine actions. The difference is
that instead of being assigned to a line of code, they are assigned to
an HTML tag that we have previously generated.
Let us suppose that the tool has already generated the following
content in the console area:
\begin{center}
\begin{minipage}{0.95\textwidth}
\begin{lstlisting}
<content format="html"/>
   <span style="color: red;" id="err1">10 errors</span> were found in sum.c
</content>
\end{lstlisting}
\end{minipage}
\end{center}
Note that the text ``10 errors'' is wrapped by a span tag with an
identifier \texttt{err1}. The OnClick action can assign a list of
commands to a click on this text as follows:
\begin{center}
\begin{minipage}{0.95\textwidth}
\begin{lstlisting}
<onclick>
  <elements> <selector value="#err1"/> </elements>
  <eicommands>
    <dialogbox boxtitle="Errors"> 
      <content format="html"> (* some text *) </content>
    </dialogbox>
  </eicommands>
</onclick>
\end{lstlisting}
\end{minipage}
\end{center}
The selectors used in the tag \lst{selector} are as in
JQuery~\cite{jquery}.

\end{document}



\end{document}